\documentclass[12pt]{article}
\usepackage{graphicx,amsmath,amssymb,psfrag}
\allowdisplaybreaks

\newcommand{\be}{\begin{equation}}
\newcommand{\ee}{\end{equation}}
\newcommand{\bea}{\begin{eqnarray}}
\newcommand{\eea}{\end{eqnarray}}

\begin{document}

\begin{titlepage}

\begin{flushright}
FERMILAB-PUB-05-385-T\\
SLAC-PUB-11468\\
{\tt hep-ph/0509090}\\[0.2cm]
September 9, 2005
\\

\end{flushright}
\vspace{0.7cm}
\begin{center}
\Large\bf 
Comment on form factor shape and extraction of $|V_{ub}|$ from $B\to\pi l \nu$ 
\end{center}

\vspace{0.8cm}
\begin{center}
{\sc Thomas Becher$^{(a)}$ and  Richard J. Hill$^{(b)}$}\\
\vspace{0.7cm}
$^{(a)}${\sl Fermi National Accelerator Laboratory,\\
P.~O.~Box 500, Batavia, IL 60510, U.S.A.} \\
\vspace{0.3cm}
$^{(b)}${\sl Stanford Linear Accelerator Center, Stanford University\\
Stanford, CA 94309, U.S.A.} 
\end{center}

\vspace{1.0cm}
\begin{abstract}
\vspace{0.2cm}
\noindent 
We point out that current experimental data for partial $B\to\pi l\nu$
branching fractions reduce the theoretical input required for a
precise extraction of $|V_{ub}|$ to the form factor normalization at a
single value of the pion energy. We show that the heavy-quark
expansion provides a bound on the form factor shape that is orders of
magnitude more stringent than conventional unitarity bounds.  We find
$|V_{ub}| = (3.7 \pm 0.2 \pm 0.1) \times [0.8/F_+(16\,{\rm GeV}^2)]$.
The first error is from the experimental branching fractions, and the
second is a conservative bound on the residual form factor shape
uncertainty, both of which will improve with additional data.
Together with current and future lattice determinations of the form
factor normalization this result gives an accurate, model independent
determination of $|V_{ub}|$. We further extract semileptonic shape
observables such as $|V_{ub}F_+(0)| = 0.92 \pm 0.11 \pm 0.03$ and show
how these observables can be used to test factorization and to
determine low-energy parameters in hadronic $B$ decays.
\end{abstract}
\vfil

\end{titlepage}

\section{Introduction \label{sec:introduction} }

Measuring the magnitude of the weak mixing matrix element $V_{ub}$ is
important for constraining the unitarity triangle and testing the
standard model of weak interactions. The exclusive determination of
$|V_{ub}|$ requires knowledge of the relevant heavy-to-light meson
form factor and in the past this has led to significant model
dependence in the result. 
First, the methods that were used to
calculate the form factor, such as light-cone sum rules, quark models
and quenched lattice calculations, all have unknown systematic
errors.  Second, 
each of these methods covers only part of the
kinematic range; to obtain the total decay rate, the results were
extrapolated using simplified parameterizations for the momentum dependence
of the form factor. 
In the past year, the situation has improved
dramatically: there are now several measurements of partial $B\to\pi
l\nu$ branching
fractions~\cite{Athar:2003yg,Abe:2004zm,Aubert:2005cd,Aubert:2005tm}
and the first results for the form factor from precision lattice
simulations with dynamical light quarks have been
presented~\cite{Okamoto:2004xg,Shigemitsu:2004ft}.  We show in this
paper that if theoretical bounds on the form factor are taken into
account, the experimental results for the partial branching fractions
determine the shape of the form factor, to the point where 
no shape information at all is required from theory.  
This reduces the
theoretical input for the determination of $|V_{ub}|$ to a
normalization of the relevant form factor, which can be taken at an
energy within the range studied with current lattice simulations.
For the first time, this allows an accurate, model independent,
determination of $|V_{ub}|$ from exclusive semileptonic decays.

Bounds on the form factor can be derived via the computation of an
appropriately chosen correlation function in perturbative QCD. By
unitarity and analyticity, the resulting ``dispersive bound''
constrains the behavior of the form factor in the semileptonic
region~\cite{Bourrely:1980gp,Boyd:1994tt,Lellouch:1995yv,Boyd:1997qw},
and may be expressed as a condition on the coefficients in a
convergent series expansion. While these bounded ``series
parameterizations'' have been around for more than twenty-five years,
many papers on the subject (in particular all experimental papers)
have instead used simple pole forms to parameterize the form factor.
In order to unify these descriptions, and to explain 
the dispersive bounds in a simple setting, 
we compare the class of series parameterizations emerging from
the conventional dispersive bound analysis to the class of ``pole
parameterizations'' introduced in \cite{Becirevic:1999kt,Hill:2005ju}.
Both representations are exact, in the sense that the true form factor is
guaranteed to be described arbitrarily well by a member of the class, 
and dispersive bounds can be established for both classes by power
counting in the heavy-quark mass. 
With the resulting constraints in place, stable fits without
truncation to a fixed number of parameters can be performed.  In fact, for the
series parameterization, we show that the bound given by simple heavy-quark power
counting is orders of magnitude more stringent than the bound based on
unitarity, thus providing much better control over the extraction of
physical observables from the data.  We also derive three new exact sum
rules for the coefficients appearing in the series representation of
the form factor.

The paper is organized as follows.  In Section~\ref{sec:params} we
review the pole and series parameterizations of the form factor.  We
list the experimental and lattice data to be used throughout the
paper, and for later comparison we determine $|V_{ub}|$ using
three-parameter truncations of these parameterizations.
Section~\ref{sec:bounds} then introduces the bounds associated with 
each parameterization, and establishes conservative estimates based on
heavy-quark power counting for the bounded quantities.
Section~\ref{sec:shape} examines the maximal precision for $|V_{ub}|$
that can be reached with present data. We show that 
with a form factor determination at intermediate $q^2$ values, within
the range studied in current lattice simulations, the experimental
uncertainty on $|V_{ub}|$ is well below 10\%, whereas a form factor
determination near maximal $q^2$ would not translate into a precise
value of $|V_{ub}|$. We introduce three shape observables,
$|V_{ub}F_+(0)|$, $F_+^\prime(0)/F_+(0)$ and $\alpha$, and discuss
their sensitivity to the exact value of the bound.  In
Section~\ref{sec:results}, having established our procedure, we
present final values for $|V_{ub}|$, and for $F_+(0)$, in terms of a
single lattice data point, $F_+(16\,{\rm GeV}^2)$.  We extract the shape
observables, which are determined by the experimental semileptonic
data alone, and show how these observables provide important
constraints on the factorization approach to hadronic $B$ decays.

\section{Form factor parameterizations and extraction of $|V_{ub}|$
  \label{sec:params}} 

\begin{figure}[t!]
\begin{center}
\psfrag{x}[B]{$q^2\, [{\rm GeV}^2]\phantom{}$}
\psfrag{y}[B]{${\small 10^4\,\tau_B\,}{d\Gamma}/{dq^2}\,\,[{\rm GeV}^{-2}]\phantom{}$}

\psfrag{a}[B]{\small \phantom{123456}CLEO \cite{Athar:2003yg}}
\psfrag{b}[B]{\small \phantom{123456}Belle \cite{Abe:2004zm}}
\psfrag{c}[B]{\small \phantom{123456}BaBar \cite{Aubert:2005cd}}
\psfrag{d}[B]{\small \phantom{123456}BaBar \cite{Aubert:2005tm}}
\includegraphics[width=0.7\textwidth]{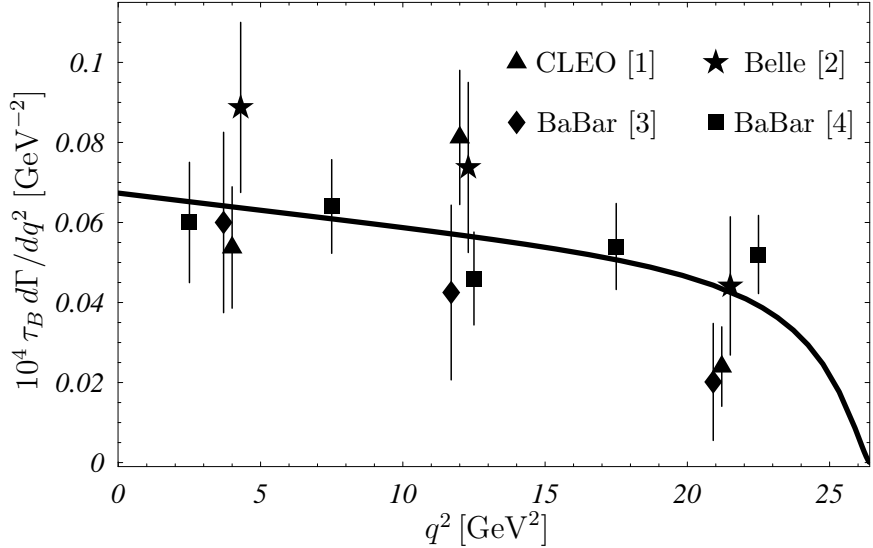}\phantom{1234}
\end{center}
  \caption{Experimental data for the partial $\bar{B}^0\rightarrow \pi^+
    \ell^-\bar{\nu}$ branching ratios and fit result shown as solid line. The
    fit results from (\ref{eq:pedestrian}) with $N=1$ and
    (\ref{eq:trunc_series}) with $k_{\rm max}=2$ are
    indistinguishable. Note that the experimental data is binned:
    \cite{Athar:2003yg,Abe:2004zm,Aubert:2005cd} give the result in
    three bins, while \cite{Aubert:2005tm} gives the result in five
    $q^2$-bins. We plot the value and error divided by the bin width
    at the average $q^2$-value in each bin. For the three-bin results,
    we have slightly shifted the points to the left and right to
    increase visibility.\label{fig:data} }
\end{figure}

Having restricted the shape of the $q^2$ spectrum, 
or equivalently, of the form factor, by experimental
measurements, the central value and errors for $|V_{ub}|$ are
determined by varying the allowed form factor over all ``reasonable''
curves that are consistent with the data, and with a normalization of
the form factor taken from theory at a given value (or multiple
values) of $q^2$.  Defining this procedure precisely requires
specifying a class of curves that contains the true form factor (to a
precision compatible with the data), and that is sufficiently rich
to describe all variations impacting the observables under study.  A
statistical analysis along standard lines then determines
central values and errors for the desired observable quantities.

A starting point to isolate such a class of curves is the dispersive
representation of the relevant form factor: 
\begin{eqnarray}\label{eq:dispersion}
  F_+(q^2) &=& {F_+(0)/(1-\alpha) \over 1 - {q^2\over m_{B^*}^2}} +
  {1\over \pi} \int_{t_+}^\infty \! dt\, {{ \rm Im} F_+(t) \over t-
    q^2-i\epsilon} \,.  
\end{eqnarray} 
Here $\alpha$ is defined by the relative
size of the contribution to $F_+(0)$ from the $B^*$ pole, and $t_\pm
\equiv (m_B \pm m_\pi)^2$.  For massless leptons, the semileptonic
region is given by $0\le q^2 \le t_-$.  Equation~(\ref{eq:dispersion})
states that, after removing the contribution of the $B^*$ pole lying
below threshold, $F_+(q^2)$ is analytic outside of a cut in the complex
$q^2$-plane extending along the real axis from $t_+$ to $\infty$,
corresponding to the production region for states with the appropriate
quantum numbers.

One class of parameterizations keeps the $B^*$ pole explicit and approximates the
remaining dispersion integral in (\ref{eq:dispersion}) by a number of effective
poles: 
\begin{equation}\label{eq:pedestrian} 
F_+(q^2) = {F_+(0)/(1-\alpha) \over 1 -
  {q^2\over m_{B^*}^2}} + \sum_{k=1}^{N} { \rho_k \over 1 - {1\over
    \gamma_k}{q^2\over m_{B^*}^2} } \,.
\end{equation}
The true form factor can be approximated to any desired accuracy by
introducing arbitrarily many, finely-spaced, effective poles.  In the
next section, we derive a bound on the magnitudes, $|\rho_k|$,  of the coefficients
of the effective poles.  This allows a meaningful
$N\to\infty$ limit, thus enabling us to investigate the behavior of
the fits when arbitrarily many parameters are included.  We find in actuality 
that current data cannot yet resolve more than one distinct effective pole
in addition to the $B^*$ pole.  Parameterizations of the above type
are widely used to fit form factors. In particular, a simplified
version of the $N=1$ case, the so-called Becirevic-Kaidalov (BK)
parameterization~\cite{Becirevic:1999kt} is used in many recent
lattice calculations and experimental studies.  As shown in
\cite{Hill:2005ju}, this two-parameter form is overly restrictive
since it enforces scaling relations which at small $q^2$ are broken by
hard gluon exchange.  The size of these hard-scattering terms, which
appear at leading order in the heavy-quark expansion, is subject to
some controversy and constraining their size is an important task. The
parameterization of the form factors should allow for their presence.

Another class of parameterizations is obtained
by expanding the form factor in a series around some $q^2=t_0$ in the semileptonic
region up to a fixed order, with the coefficients of this
expansion as the fit parameters.  
The convergence of this simple expansion is
very poor due to the presence of the nearby singularities at
$q^2=m_{B^*}^2$ and $q^2=t_+$.   However, 
an improved series expansion of the form factor that converges in the
entire cut $q^2$-plane is obtained after a change of variables that
maps this region onto the unit disc $|z|<1$. In terms of the new
variable, $F_+$ has an expansion
\begin{equation}
  \label{eq:series}
  F_+(q^2) = { 1\over P(q^2)\phi(q^2,t_0)} \sum_{k=0}^\infty a_k(t_0)\,
  [z(q^2,t_0)]^k \,, \quad z(q^2,t_0) = {\sqrt{t_+ - q^2} - \sqrt{t_+-
      t_0} \over \sqrt{t_+ - q^2} + \sqrt{t_+ - t_0}} \,, 
\end{equation} 
with real coefficients $a_k$. The variable $z(q^2,t_0)$ maps the
interval $-\infty< q^2< t_+$ onto the line segment $-1<z<1$, with the
free parameter $t_0 \in (-\infty, t_+)$ corresponding to the value of
$q^2$ mapping onto $z=0$.  Points immediately above (below) the
$q^2$-cut are mapped onto the lower (upper) half-circle $|z|=1$.  The
function $P(q^2) \equiv z(q^2, m_{B^*}^2)$ accounts for the pole in
$F_+(q^2)$ at $q^2=m_{B^*}^2$, while $\phi(q^2)$ is any function
analytic outside of the cut. It is interesting to note that this
reorganization succeeds in turning a large recoil parameter, $(v\cdot
v^\prime)_{\rm max}-1 \approx 18$, into a small expansion parameter.  For example, for $t_0=0$ the variable $z$ is negative
throughout the semileptonic region and 
\begin{equation}
  |z|_{\rm max} = \frac{
    \sqrt{(v\cdot v^\prime)_{\rm max}+1 } - \sqrt{2} }{
    \sqrt{(v\cdot v^\prime)_{\rm max}+1 } + \sqrt{2} }\approx 0.5\, ,
\end{equation} where $v$
and $v^\prime$ are the velocities of the $B$ and $\pi$ mesons.  The
same size, but for positive $z$ is obtained for $t_0=t_-$. By choosing
the intermediate value $t_0=t_+( 1 - \sqrt{1-t_-/t_+})$, the expansion
parameter can be made as small as $|z|_{\rm max} \approx 0.3$.  A
second class of parameterizations is obtained by a truncation of the
above series:
\begin{equation}
  \label{eq:trunc_series}
  F_+(q^2) = { 1\over P(q^2)\phi(q^2,t_0)} \sum_{k=0}^{k_{\rm max}} a_k(t_0) \,
  [z(q^2,t_0)]^k \,.
\end{equation}
As discussed in the next section, it is conventional to take 
\begin{equation}\label{eq:phi} 
  \phi(q^2,t_0) = \left( \pi m_b^2 \over 3 \right)^{1/2}
  \left( z(q^2,0)\over -q^2 \right)^{5/2}\left(z(q^2,t_0)\over t_0-q^2
  \right)^{-1/2} \left( z(q^2,t_-)\over t_- - q^2 \right)^{-3/4}{(t_+-q^2)
    \over (t_+ - t_0)^{1/4}} \,.    
\end{equation}
With this choice, a bound $\sum_k a_k^2 \lesssim 1$ is obtained by perturbative methods.~%
\footnote{
  For different $t_0$,
  the expansion parameters, $z\equiv z(t,t_0)$ and $z^\prime\equiv z(t,t_0^\prime)$,     
  and expansion coefficients, $a_k\equiv a_k(t_0)$ and $a_k^\prime \equiv a_k(t_0^\prime)$, 
  are related by the M\"{o}bius transformation: 
  \begin{equation}\nonumber
    z^\prime = { z(t_0, t_0^\prime) + z \over 1 + z(t_0, t_0^\prime) z } \,, \quad 
    \sqrt{1-z^2} \sum_{k=0}^\infty a_k z^k 
    = \sqrt{1-{z^\prime}^2} \sum_{k=0}^\infty a_k^\prime {z^\prime}^k \,. 
  \end{equation}
  It is easily verified that the sum of squares of coefficients is invariant 
  under such a transformation, 
  $\sum_k a_k^2 = \sum_k {a_k^\prime}^2$, as guaranteed by the construction of 
  $\phi$, see (\ref{eq:bound}). 
} 
Together with the restriction $|z|<1$, this allows a meaningful $k_{\rm max}\to \infty$ 
limit.   In actuality, we find that current data can only resolve the first three terms
in the series (\ref{eq:trunc_series}). 

\begin{figure}[t!]
\begin{center}
\psfrag{x}[B]{\small$F_+(16\,{\rm GeV}^2)$}
\psfrag{y}[B]{\small$|V_{ub}| \times 10^{3}$}
\begin{tabular}{cc}
\includegraphics*[height=0.4\textwidth]{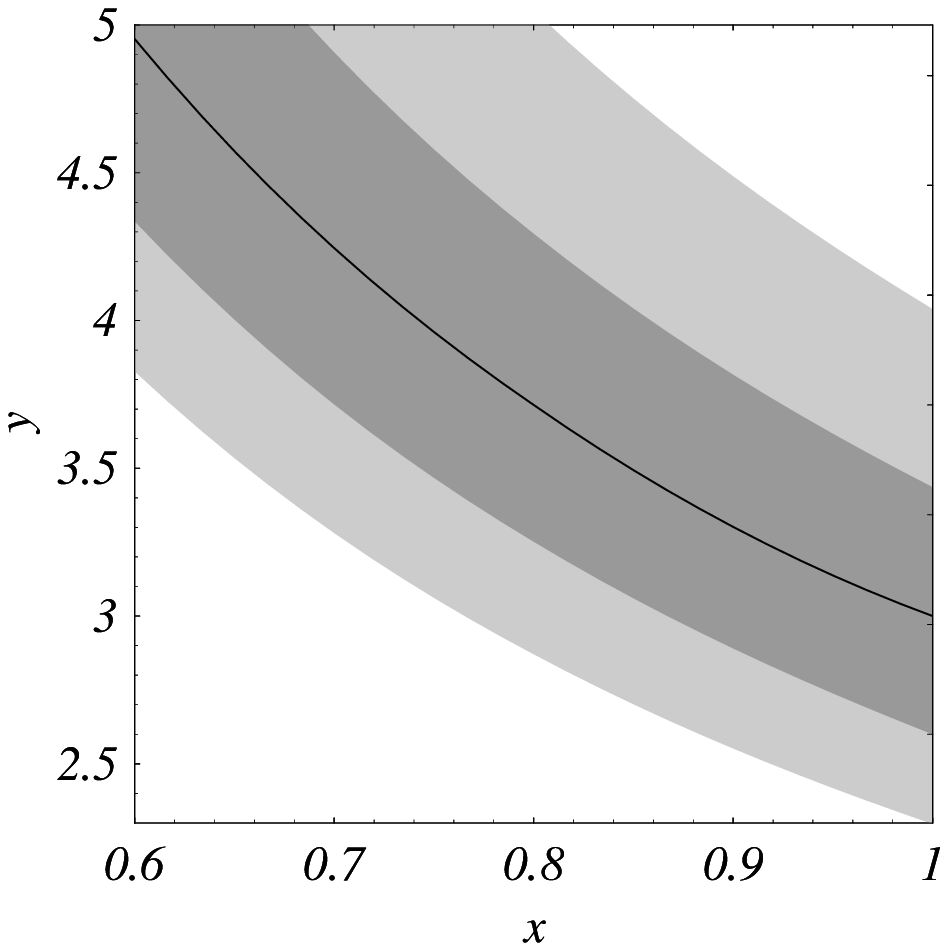} &
\psfrag{x}[B]{\small $\Delta F_+/F_+ 
$}
\raisebox{-0.012\textwidth}{\includegraphics*[height=0.424\textwidth]{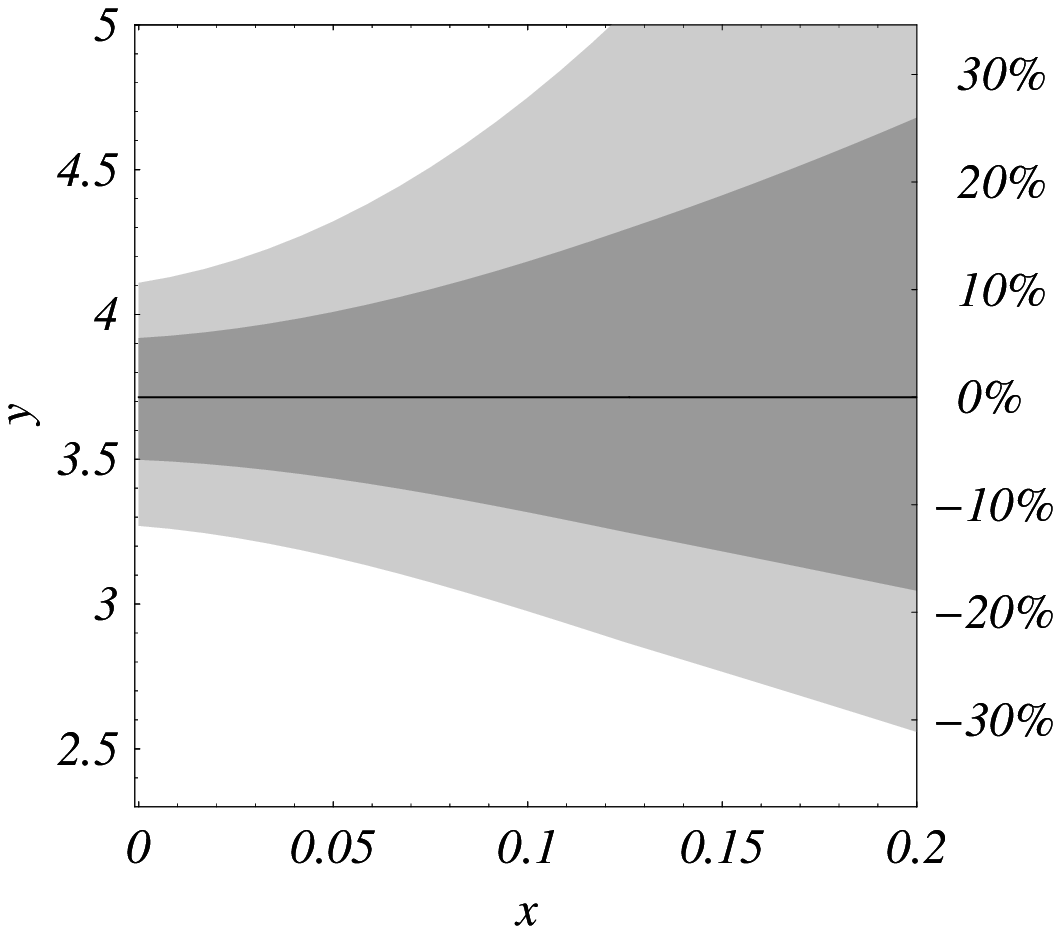}}
\end{tabular}
\end{center}
\caption{$68\%$ (dark) and $95\%$ (light) confidence limits 
  for $|V_{ub}|$ determined by fitting the parameterizations
  (\ref{eq:pedestrian}) or (\ref{eq:trunc_series}) to experimental
  data in \cite{Athar:2003yg}, \cite{Abe:2004zm}, \cite{Aubert:2005cd}
  and \cite{Aubert:2005tm}, with the single lattice data point  
  $F_+(16\,{\rm GeV}^2) = 0.8\pm 0.1$. 
  Results from (\ref{eq:pedestrian}) and
  (\ref{eq:trunc_series}) are indistinguishable.  
  The plot on the right shows
  $|V_{ub}|$ for fixed $F_+(16\,{\rm GeV}^2)=0.8$ as a function of the
  relative uncertainty on the form factor.   }
\label{fig:vub}
\end{figure}

Figure~\ref{fig:data} shows the available experimental data on the
partial branching fraction  $d\Gamma(\bar B^0\rightarrow \pi^+ \ell^- \bar{\nu})/dq^2$. 
The CLEO~\cite{Athar:2003yg}, 
Belle~\cite{Abe:2004zm} and BaBar~\cite{Aubert:2005tm} 
collaborations 
have measured this branching fraction in three separate 
$q^2$-bins and BaBar~\cite{Aubert:2005cd} has presented a measurement
using five $q^2$-bins. 
The correlation matrix is included in our fits for the data in
\cite{Athar:2003yg}.  For the remaining data, $q^2$-bins are taken as
uncorrelated.
In order to extract $|V_{ub}|$ we also need the normalization of the form factor. 
The $B\rightarrow \pi$ vector form factors have recently been determined by the
Fermilab Lattice~\cite{Okamoto:2004xg} and by the 
HPQCD~\cite{Shigemitsu:2004ft}
collaborations in lattice simulations with dynamical fermions. The
preliminary results of these calculations give 
$F_+(16\,{\rm  GeV}^2)=0.81\pm 0.11$~\cite{Okamoto:2004xg} 
and $F_+(16\,{\rm  GeV}^2)= 0.73\pm 0.10$~\cite{Shigemitsu:2004ft}.~%
\footnote{ The parameterization (\ref{eq:pedestrian}) with $N=1$ has
  been used to interpolate to the common $q^2$-point, and for
  definiteness the errors are taken as those from the nearest points:
  $q^2=15.87\,{\rm GeV}^2$~\cite{Okamoto:2004xg}, with statistical and
  systematic errors added in quadrature,  and $q^2=16.28\,{\rm
    GeV}^2$~\cite{Shigemitsu:2004ft}. 
} 
Although the lattice calculations give the form factor at several different
$q^2$-values, the
correlations between different points are not available and it is
difficult to quantify the uncertainty on the shape. 
Anticipating the analysis of Section~\ref{sec:shape}, where we determine the 
range of $q^2$ that best exploits the experimental shape information,  
we use $F_+(16\,{\rm  GeV}^2)=0.8\pm 0.1$ as our default value for the form factor
normalization.   
Note that we have avoided any theoretical biases concerning the form factor shape. 
Performing a $\chi^2$ fit 
yields $|V_{ub}| = 3.7^{+0.6}_{-0.5}\times
10^{-3}$ for both the parameterization (\ref{eq:pedestrian}) with
$N=1$, and (\ref{eq:trunc_series}) with $k_{\rm max}=2$.

Figure~\ref{fig:vub} shows the $68\%$ and $95\%$
confidence limits for $|V_{ub}|$ as a function of the value and
uncertainty of the form factor at $q^2=16\,{\rm GeV}^2$. 
The form factor normalization is the dominant error in
the determination of $|V_{ub}|$; if the quantity $F_+(16\,{\rm GeV}^2)$ 
would be known
exactly, the uncertainty on $|V_{ub}|$ would drop to approximately $6\%$.
The quality of the fit is equally good for both parameterizations, with
$\chi^2=12.0$.~%
\footnote{ Note that all three-bin measurements determine the same
  observable quantities.  The minimal $\chi^2$ obtained from the
  three-bin measurements is $5.0$ for $9-3$ degrees of freedom.  This
  value measures the (good) agreement between the three-bin
  measurements, and should be subtracted from the total in order to
  obtain a measure of agreement between the data and the
  parameterizations.  The resulting quality of our fit is good:
  $12.0-5.0$ for $9-4$ degrees of freedom.  } The extracted value of
$|V_{ub}|$ is insensitive to the choice of the free parameter $t_0$.
Setting $\phi(q^2)=1$ in (\ref{eq:trunc_series}) also has negligible
impact, and similarly adding more lattice input points does not
substantially change the result if the dominant lattice errors are
correlated.  The effect of allowing additional terms in the
parameterizations (\ref{eq:pedestrian}) and (\ref{eq:trunc_series}) is
investigated in the following sections.  We will find that the result
for the value and uncertainty of $|V_{ub}|$ from the simple
parameterizations used in this section is not appreciably altered if
additional terms are included.

\section{Form factor bounds\label{sec:bounds}} 

To make a fully rigorous determination of $|V_{ub}|$, the truncation
to the three-parameter classes of curves considered in the previous
section requires justification.  For instance, if the neglected terms
in (\ref{eq:pedestrian}) or (\ref{eq:trunc_series}) conspired to
produce a sharp peak in the form factor at precisely the value of the
lattice input point, then the integrated rate would be overestimated,
and the value of $|V_{ub}|$ underestimated.  To prevent this from
happening requires some bound on the perversity of allowed form factor
shapes.
In practice, we would like to ensure that our extraction of physical 
observables is ``model-independent'' by allowing for arbitrarily many 
parameters, i.e., taking $N\to\infty$ in (\ref{eq:pedestrian}) and
$k_{\rm max}\to\infty$ in (\ref{eq:trunc_series}).  
Retaining predictive power then demands that a bound be enforced on the
parameters appearing in these expansions. 

To bound the coefficients $\rho_k$ in the expansion (\ref{eq:pedestrian}),
we introduce a decomposition of the integration region,
$t_+ \le t_1 < \dots < t_{N+1} < \infty$,  
and define 
\begin{equation}\label{eq:rho}
  \rho_k \equiv {1\over \pi}\int_{t_{k}}^{t_{k+1}}\!{dt\over t}\, {\rm Im} F_+(t) \,, \quad
  \gamma_k\equiv { t_k \over  m_{B^*}^2}  \,.  
\end{equation}
Since $F_+(t)\sim t^{-1}$ at large $t$, it follows that 
\begin{equation} \label{eq:bound_rho} 
  \sum_k |\rho_k| \le {1\over\pi}\int_{t_+}^\infty {dt\over t} |F_+(t)| \equiv R \,,
\end{equation}
and this is the desired bound.  The integral in
(\ref{eq:bound_rho}) is dominated by states with $t-t_+\sim
m_b\Lambda$, where $F_+\sim m_b^{1/2}$, so that the quantity $R$ is
parametrically of order $(\Lambda/m_b)^{1/2}$, with $\Lambda$ a
hadronic scale.  To be sure that the bound deserves the
model-independent moniker, one should use a very conservative estimate. 
In our fits we will use $R\le \sqrt{10}$ and $R \le 10$, i.e., 
we allow for an addition factor of $100$ or $1000$ beyond the  
dimensional estimate $R^2 \sim \Lambda/m_b \sim 0.1$.    

The coefficients $a_k$ in the expansion (\ref{eq:trunc_series}) can be 
bounded by requiring that the 
production rate of $B\pi$ states, described by the analytically continued form factor, 
does not overwhelm the production rate of {\it all} 
states coupling to the current of interest 
(in this case, the vector current $\bar{u} \gamma^\mu b$).  
The latter rate is computable in perturbative QCD using the operator product expansion
(for a pedagogical discussion, see e.g. \cite{Boyd:1997qw}). 
The function $\phi$ in
(\ref{eq:phi}) was chosen such that the fractional contribution of
$B\pi$ states to this rate is given at leading order by
\begin{equation}\label{eq:bound} 
  \sum_{k=0}^{\infty} a_k^2 
  = {1\over 2\pi i}\oint {dz\over z}
  |\phi(z) P(z) F_+(z)|^2 = 
  {m_b^2\over 3} \int_{t_+}^\infty\! {dt \over t^5}
  [(t-t_+)(t-t_-)]^{3/2} |F_+(t)|^2 
  \equiv A \,.  
\end{equation}
In the heavy quark limit, the leading contributions to the integral
$A$ in (\ref{eq:bound}) are of order $(\Lambda/m_b)^3$ and arise from
two regions: the region close to threshold, $t-t_+\sim m_b\Lambda$, 
where the pion has energy $E\sim \Lambda$ and the form factor scales as
$F_+\sim m_b^{1/2}$; 
and the region $t-t_+ \sim m_b^2$, where $E\sim m_b$ 
and $F_+\sim m_b^{-3/2}$ (for a discussion of the form factor scalings, see 
\cite{Hill:2005ju}). 
%
%
The region of very high
energies $t\gg m_b^2$, where $F_+\sim 1/t$, gives a subleading
contribution.

Since, by definition, the fraction is smaller than unity, it is
conventional to take the loose bound $A \le 1$, which does not make
use of scaling behavior in the heavy-quark limit.  Clearly this bound
leaves much room for improvement; from its scaling behavior, we expect
$A$ to be on the order of a few permille.  This implies that
higher-order perturbative and power corrections in the operator
product analysis introduce negligible error, as noticed in
\cite{Arnesen:2005ez}.  It is also easy to see that the dispersive
bounds by themselves do not impose tight constraints on the form
factor shape.  Since the scale of the coefficients is set by $a_0\sim
m_b^{-3/2}$, even with the optimal choice $|z|_{\rm max}\approx 0.3$,
the dispersive bounds allow the relative size of higher-order terms in
the series, $|a_k z^k/ a_0|$, to be of order unity up to $k\approx 4$,
and to contribute significantly for even higher $k$.  This situation
for heavy-to-light decays such as $B\to\pi$ contrasts with that for
heavy-heavy decays such as $B\to D$~\cite{Boyd:1995sq,Caprini:1997mu},
where the bound is parametrically of order unity (counting $m_c \sim
m_b\gg \Lambda$).  For this case, the scale of the coefficients is set
by $a_0 \sim m_b^0$, and with $|z|_{\rm max}\approx 0.06$ the bound
ensures that only the first few terms in the series are required for
percent accuracy.

To put the dispersive bounds in perspective, it may be useful to
emphasize that establishing an order-of-magnitude bound on any
integral of the form $\int_{t_+}^\infty\!dt\, k(t) |F_+(t)|^2$ for
some $k(t)$ would yield an equally valid, bounded, parameterization,
with a new $\phi(t)$ constructed from $k(t)$ as in (\ref{eq:phi}) and
(\ref{eq:bound}).  Similarly, bounded pole parameterizations
(\ref{eq:pedestrian}) are obtained by establishing an
order-of-magnitude bound on any integral of the form
$\int_{t_+}^\infty\!dt\, k(t) |F_+(t)|$ for some $k(t)$, as in
(\ref{eq:rho}) and (\ref{eq:bound_rho}).  Focusing attention on the special case of
(\ref{eq:phi}) and (\ref{eq:bound}) is justified only to the extent
that the bound (\ref{eq:bound}) is sufficiently restrictive, and to
the extent that similar or tighter bounds cannot be conservatively
estimated by other means.

It is interesting to note that the two bounds are not equivalent. The
bound $\sum_k |\rho_k| \le R < \infty $ uses the fact that the asymptotic form factor
can be evaluated in perturbation theory, where the scaling $F_+(t) \sim t^{-1}$ 
is found at large $t$. 
This condition is not automatically
satisfied by the series parameterization (\ref{eq:trunc_series}), which as seen 
from (\ref{eq:bound}) requires only $F_+(t) \lesssim t^{1/2}$ at large $t$. 
Imposing the proper large-$t$ behavior yields the sum rules%
\begin{equation}\label{eq:sumrules}
  \frac{d^n}{dz^n}P(z)\phi(z)F(z)\bigg{|}_{z=1}=0\;
  \quad \leftrightarrow \quad
  \sum_{k=0}^\infty\, k^n\, a_k z^k \bigg{|}_{z\to 1} =0\,, \quad n=0,1,2.
\end{equation}
To our knowledge, the above sum rules have not been discussed in the
literature.  On the other hand, all pole parameterizations ``violate''
the bound $\sum_{k=0}^\infty a_k^2 \equiv A < \infty$ for the simple reason that the
integral in (\ref{eq:bound}) is not well defined for these
parameterizations, because $F_+(t)$ has poles on the integration
contour.

The bounds discussed here are associated with the behavior of the form
factor above threshold. Since we are interested in the form factor in
the semileptonic region, these higher-energy properties are useful
only to the extent that they can help to constrain the form factor in
this region. Incorrect high-energy behavior therefore does not imply
that a given parameterization cannot be used to describe
low-energy data.  For instance, the effective poles in
(\ref{eq:pedestrian}) could be smeared into finite-width effective
resonances in order to make the integral in (\ref{eq:bound}) converge;
however, the semileptonic data is very insensitive to such
fine-grained detail, and this modification has a very minor impact on
the fits.  Similarly, unless the bound (\ref{eq:bound}) is close to
being saturated, the coefficients $a_k$ for moderately large $k$ in
the series parameterization (\ref{eq:trunc_series}) can be tuned to
satisfy the sum rules (\ref{eq:sumrules}), or equivalently, to make
the integral in (\ref{eq:bound_rho}) converge.  However, the
semileptonic data becomes insensitive to terms $z^k$ for large $k$,
and again such a modification has little impact on the fits.  Thus,
while at some level the bound (\ref{eq:bound_rho}) will constrain the
parameters in the series parameterization (\ref{eq:trunc_series}), and
the bound (\ref{eq:bound}) will constrain the parameters in the pole
parameterization (\ref{eq:pedestrian}), we restrict attention to the
constraints imposed by (\ref{eq:bound_rho}) on the pole
parameterization, and by (\ref{eq:bound}) on the series
parameterization.

\section{Parameterization uncertainty and shape observables\label{sec:shape}}

\begin{figure}[t]
\begin{center}
\psfrag{q}{$q^2\,[{\rm GeV}^2]$}
\psfrag{y}[B]{${ 10^3} |V_{ub}|$}
\includegraphics[width=0.8\textwidth]{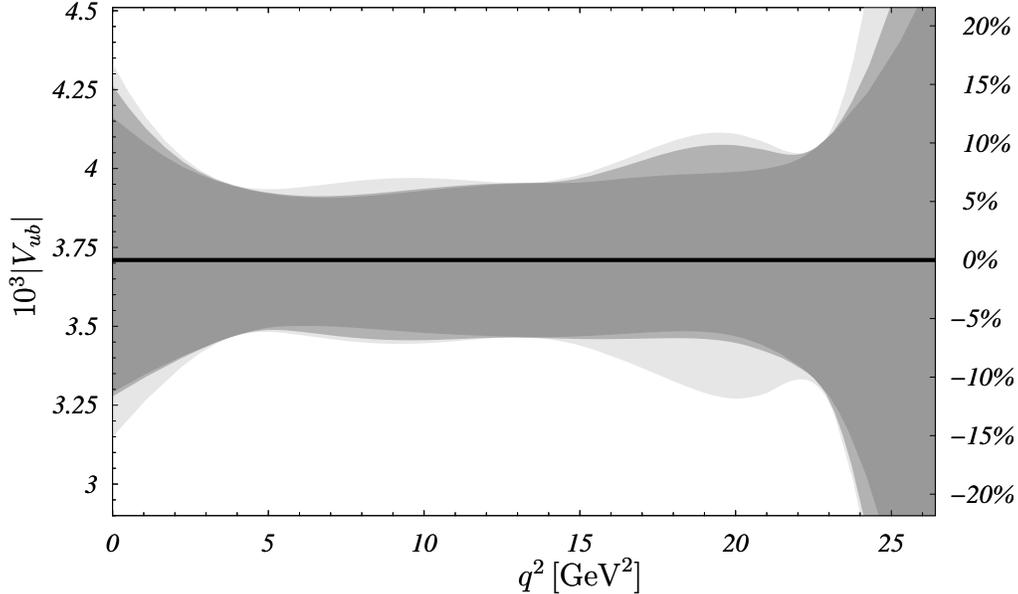}
\end{center}
  \caption{$\Delta \chi^2=1$ region for $|V_{ub}|$ for an
    infinitely precise form factor determination at a single
    $q^2$-value. The plot assumes that the form factor yields the 
    central value $|V_{ub}|=3.7\times 10^{-3}$.  The
    darkest band is obtained for $\sum_k a_k^2<0.01$, while the two lighter
    bands correspond to $\sum_k a_k^2<0.1$ and $\sum_k a_k^2<1$.
    \label{fig:q2val}  
  }
\end{figure}

With the bounds in place, it is straightforward to generalize the 
fits in Section~\ref{sec:params} to include arbitrarily many parameters. 
Imposing the very conservative bound $\sum_k |\rho_k| < 10$, 
we observe that 
additional
poles in the class of parameterizations (\ref{eq:pedestrian})
have essentially no impact on the central value and errors for
$|V_{ub}|$.
Similarly, using the very conservative 
bound $\sum_{k} a_k^2 < 1$ in (\ref{eq:trunc_series}), 
we find that the inclusion of higher order
terms beyond $k_{\rm max}=2$ has negligible impact on $|V_{ub}|$. 
The errors are dominated by the lattice input point, 
and both the central value and errors are not changed significantly from
the  $N=1$ or $k_{\rm max}=2$ fits in Section~\ref{sec:params}. 

In order to isolate the uncertainty on the form factor shape 
inherent to the present data, we show in Figure~\ref{fig:q2val} the minimum attainable 
error on $|V_{ub}|$, 
assuming exact knowledge of the form factor at one 
$q^2$-value.  Results are shown for the parameterization (\ref{eq:trunc_series}), 
using various bounds $\sum_k a_k^2 < 0.01$, $0.1$ and $1$.    
As the figure illustrates, 
points in the intermediate range of $q^2$ lead 
to the smallest uncertainty on $|V_{ub}|$,  
and for these points, the $|V_{ub}|$ extraction is not very 
sensitive to even the order of magnitude of the chosen bound, 
with the minimum error varying from approximately $6\%$ to approximately 
$8\%$ as the bound is relaxed from $0.01$ to $1$. 
It should be noted that a 
better understanding of correlations in the experimental data would 
be necessary when probing this level of precision. 
The curves in Figure~\ref{fig:q2val} are also indicative of the impact of 
additional theory inputs.  Performing the fits with data points at different 
$q^2$-values in addition to the default $F_+(16\,{\rm GeV}^2)$  
 shows that a point at $q^2=0$ would require $\lesssim 10\%$ error 
to significantly decrease the error on $|V_{ub}|$, while
even exact knowledge of the form factor at $q^2=t_-$ has almost no impact.   

In the remainder of 
this section we consider observables which are more sensitive to
the shape of the form factor and investigate the role played by the bounds in
these cases.  In particular, we extract the form factor and its first derivative at $q^2=0$, 
as well as the residue at the $B^*$ pole, which is directly related to the parameter
$\alpha$, as in (\ref{eq:dispersion}). 
These quantities are interesting
in their own right. The form factor at zero momentum transfer, 
normalized as $|V_{ub}|F_+(0)$,  
is an important input for the evaluation of factorization theorems for
charmless two-body decays such as $B\rightarrow \pi\pi$.  
The derivative of the form factor at $q^2=0$, 
conveniently normalized as $(m_B^2-m_\pi^2)F_+^\prime(0)/F_+(0)$,
determines the quantity $\delta$
measuring the ratio of hard-scattering to soft-overlap terms in
the form factor~\cite{Hill:2005ju}.
Finally, the value of $(1-\alpha)^{-1}$ is proportional to the coupling
constant $g_{B^*B\pi}$. 
The observable quantities  $|V_{ub}|F_+(0)$, $F_+^\prime(0)/F_+(0)$ and $\alpha$ 
are independent of the form factor
normalization, and hence are determined solely by the experimental
data. 

\begin{table}[t!]
\begin{center}
\begin{tabular}{|c|ccc|ccc|}\hline
bound & & $\sum_k |\rho_k|=10$ & & &$\sum_k |\rho_k|=\sqrt{10}$ & \\[3pt] \hline 
$N$ & $1$ & $2$ & $3$ & $1$ & $2$ & $3$ \\
$\sum_k |\rho_k|$ & $1.02$ & $1.36$ & $10$ & $1.02$ & $1.36$ & $\sqrt{10}$ \\
$\chi^2$ & $11.97$ & $11.96$ & $11.58$ & $11.97$ & $11.96$ & $11.80$ \\[3pt] \hline  
&&&&&&\\[-15pt]
$10^3 |V_{ub}|F_+(0)$ & $0.93_{-0.09}^{+0.06}$ & $0.93_{-0.09}^{+0.11}$ & $0.87_{-0.12}^{+0.14}$ & \
$0.93_{-0.09}^{+0.06}$ & $0.93_{-0.09}^{+0.10}$ & $0.91_{-0.10}^{+0.11}$ \\[5pt]
$\frac{(m_B^2-m_\pi^2)F_+^\prime(0)}{F_+(0)}$ 
& $1.3_{-0.1}^{+0.4}$ & $1.3_{-0.7}^{+0.4}$ & $2.0_{-1.2}^{+0.9}$ & \
$1.3_{-0.1}^{+0.4}$ & $1.3_{-0.6}^{+0.4}$ & $1.5_{-0.8}^{+0.6}$ \\[5pt]
$ (1-\alpha)^{-1} $  & $5_{-3}^{+5}$ & $6_{-5}^{+6}$ & $6_{-15}^{+20}$ 
& $5_{-3}^{+5}$ \
& $6_{-5}^{+6}$ & $6_{-8}^{+6}$ \\ \hline
\end{tabular} 
\end{center}
\caption{Fit results for form factor shape parameters
  using the pole parameterization (\ref{eq:pedestrian}).\label{tab:fitsN}
}
\end{table}

In Tables \ref{tab:fitsN}  and \ref{tab:fitskmax}, we show how the
results for the shape observables  change when additional parameters are
included.  
For the pole parameterization (\ref{eq:pedestrian})
we perform fits with
$N=1$, $2$ and $3$ poles in addition to the $B^*$ pole.    
(The case $N=1$ was studied in \cite{Hill:2005ju}.) 
To help stabilize the fits, we impose a minimum spacing of the poles
$\gamma_{k+1}-\gamma_k > 1/N$, and a maximum pole position, $\gamma_k
< N + 1$. For the polynomial parameterization (\ref{eq:trunc_series}),
we set $k_{\rm max}=2$, $3$ and $4$.  We perform each of the fits with
two different bounds --- a very loose model-independent bound, and a more
stringent bound that relies on the scaling behavior of the bounded
quantity in the heavy-quark limit.  Given a value of the bound, a
central value and errors are determined by taking the limit of large
$N$ in (\ref{eq:pedestrian}), or large $k_{\rm max}$ in
(\ref{eq:trunc_series}).  The sequence converges once the size of the
neglected terms is constrained by the bound to lie below the
sensitivity of the chosen observable.

The quantities $|V_{ub}|F_+(0)$, $F_+^\prime(0)/F_+(0)$ and $\alpha$
exhibit different sensitivities to the bounds.  This is to be
expected, since sharp bends in the fitted curve at the endpoints
allowed by the additional terms can have strong effects on the slope,
or on the residue of the $B^*$ pole, but are not constrained tightly
by the data.  Imposing only very loose bounds therefore leads to large
uncertainties for these quantities.

It is instructive to examine the relation between 
observables and expansion coefficients.  
At $t_0=0$ the quantities $f(0)$, 
$\alpha$, $\beta$ and $\delta$ studied in \cite{Hill:2005ju} are related to the 
coefficients $a_k$ by 
\begin{align}\label{eq:id} 
  f(0) \equiv F_+(0)
  &= 
  {16 a_0\over\hat{m}_b} 
  \left(3\over\pi \right)^{1/2}{(1+\hat{m}_\pi)^{5/2}\over (1+ \sqrt{\hat{m}_\pi})^3}
  {1 + \hat{m}_\pi + \hat{\Delta} \over 1 + \hat{m}_\pi - \hat{\Delta} } \,, 
  \nonumber\\ 
  {1+\beta^{-1}-\delta} \equiv {m_B^2 -m_\pi^2 \over F_+(0)}\left.{dF_+\over dq^2}\right|_{q^2=0} 
  &=
  {-a_1\over 4a_0}  {1-\hat{m}_\pi\over 1+\hat{m}_\pi} + 
  {3\over 4} {1-\sqrt{\hat{m}_\pi}\over 1+ \sqrt{\hat{m}_\pi}} 
  + {\hat{\Delta} (1-\hat{m}_\pi)\over (1+\hat{m}_\pi)^2 - \hat{\Delta}^2}  \,, 
  \\
  {(1-\alpha)^{-1}} 
  &=
  { (1+\hat{m}_\pi + \hat{\Delta})^2 (1+\sqrt{\hat{m}_\pi})^3
    \over 4(1+\hat{m}_\pi)^2 (\hat{\Delta} + 2\sqrt{\hat{m}_\pi})^{3/2} }
  \sum_{k=0}^{\infty} {a_k\over a_0}\!\! 
  \left((-1)  { 1+\hat{m}_\pi - \hat{\Delta} \over 1+\hat{m}_\pi + \hat{\Delta} } \right)^k ,\nonumber
\end{align}
where $\Delta^2\equiv (m_B+m_\pi)^2 - m_{B^*}^2 $, and hats denote
quantities in units of $m_B$. The heavy-quark scaling laws for $f(0)$
and $1+\beta^{-1}-\delta$ are special cases of the general law $a_k
\sim m_b^{-3/2}$, obtained by taking $k$ derivatives in
(\ref{eq:series}), and noticing that 
$d^n F_+/d {({\hat q}^2})^n|_{q^2=0} \sim m_b^{-3/2}$ when scaling violations are neglected. Similarly, the scaling law for
$(1-\alpha)^{-1}$ translates into the behavior $\hat{\Delta}^{-1/2}
\sim m_b^{1/4}$ for the sum appearing in the last line of
(\ref{eq:id}).

\begin{table}[t!]
\begin{center}
\begin{tabular}{|c|ccc|ccc|}\hline
  bound & & $\sum_k a_k^2<1$ & & &$\sum_k
  a_k^2<0.01$& \\[3pt] \hline 
  $k_{\rm max}$ & 2 & 3 &4 & 2 & 3 & 4 \\
  $\sum_k a_k^2$ &  $0.003$ & $0.3$ & $1$ & $0.003$ & $0.01$ & $0.01$ \\
  $\chi^2$ & $12.0$ & $11.7$ & $11.7$ & $12.0$ & $11.9$ & $11.9$ \\[3pt]\hline
  &&&&&&  \\[-8pt]
  $10^3 |V_{ub}|F_+(0)$ &  $0.93_{-0.10}^{+0.10}$ & $0.87_{-0.15}^{+0.15}$
  & $0.87_{-0.14}^{+0.14}$ &  $0.93_{-0.10}^{+0.10}$ & $0.92_{-0.10}^{+0.11}$ & $0.92_{-0.10}^{+0.11}$ \\[5pt]
  $\frac{(m_B^2-m_\pi^2)F_+^\prime(0)}{F_+(0)}$ &  $1.3_{-0.5}^{+0.6}$ & $2.0_{-1.4}^{+1.4}$ & $2.0_{-1.4}^{+1.4}$ &
  $1.3_{-0.4}^{+0.6}$ & $1.4_{-0.6}^{+0.6}$ & $1.5_{-0.6}^{+0.6}$  \\[5pt]
  $(1-\alpha)^{-1}$  & $6_{-2}^{+2}$ & $13_{-14}^{+8}$ & $9_{-17}^{+20}$ & 
  $6_{-2}^{+2}$ & $7_{-5}^{+2}$ & $8_{-6}^{+2}$
  \\ \hline
\end{tabular} 
\end{center}
\caption{Fit results for form factor shape parameters using the
  series parameterization (\ref{eq:trunc_series}) with
  $t_0=0$.\label{tab:fitskmax}} 
\end{table} 

\section{Results and discussion\label{sec:results}}

In order to extract the most precise value of 
$|V_{ub}|$, it
is important to make full use of the existing experimental data for
$B\to\pi l\nu$ that determines the form factor shape.  To emphasize
this point, the analysis was done here using no shape information at
all from theory, but only a normalization at one $q^2$-point.
Our results make it clear that the limiting factor in the determination
of $|V_{ub}|$ is currently the form factor normalization, with very small 
uncertainty associated with the form factor shape. 
Similar conclusions are implicit in other recent works.  For example,
in \cite{Arnesen:2005ez} the reduction in error compared to methods 
employing only 
total experimental branching fractions is due almost entirely to the
inclusion of shape information from experiment, and not to the
inclusion of additional theory input points.  
In \cite{Ball:2005tb}, experimental data is 
combined with simple parameterizations of the form factor shape to 
constrain the hadronic input parameters appearing in sum rule estimates of the
form factor. 
In contrast to these and other previous works, we have avoided any theoretical 
biases concerning the form factor shape.  
  
In practical terms, the parameterizations (\ref{eq:pedestrian}), with
$N=1$, and (\ref{eq:trunc_series}), with $k_{\rm max}=2$, are
sufficient for describing the current generation of semileptonic data, in the 
sense that the addition of more parameters does not significantly improve the 
fits.
To provide rigorous error estimates it is necessary to allow for arbitrarily many 
additional parameters within the dispersive bounds (\ref{eq:bound_rho}) and (\ref{eq:bound}). 
For ``global'' quantities like $|V_{ub}|$ it is possible to show by 
imposing only the very loose bounds 
$\sum_k|\rho_k|<10$ in (\ref{eq:pedestrian}), 
or $\sum_k a_k^2< 1$ 
in (\ref{eq:trunc_series}) that the extracted values are
actually insensitive to the addition of more parameters.  
With a single lattice input value $F_+(16\,{\rm GeV}^2)=0.8\pm 0.1$, we find
\begin{equation}
\begin{aligned}\label{eq:norm}
  |V_{ub}| 
  &= 3.7 \pm 0.2\, ^{+0.6}_{-0.4} \pm 0.1
  &&= \left( 3.7 \pm 0.2 \right) \times { 0.8 \over F_+(16\,{\rm GeV}^2)} \,, \\
  F_+(0)  
  &= 0.25\pm{0.04}\pm{0.03} \pm 0.01  
  &&= \left( 0.25 \pm 0.04 \right) \times { F_+(16\,{\rm GeV}^2) \over 0.8 } 
  \,. 
\end{aligned}
\end{equation}
The first error is experimental, the second is theoretical from the lattice input, 
and the third is due to the uncertainty in the form factor shape.  
For definiteness, the central values in (\ref{eq:norm}) are obtained using the
parameterization (\ref{eq:trunc_series}) with $\sum_k a_k^2 < 0.01$, 
and the third error 
is very conservatively estimated by adding the maximum variation of the boundaries of the
$1\sigma$ interval induced by relaxing the bound to $\sum_k a_k^2 < 1$.   

For less global quantities, like the slope of the form factor
at $q^2=0$, the very loose bounds (\ref{eq:bound_rho}) and 
(\ref{eq:bound}) are not sufficient to tightly
constrain the impact of arbitrarily many additional parameters.
In this case we adopt more realistic estimates for the bounds, and find
\begin{align}\label{eq:shape}
  10^3|V_{ub}F_+(0)| 
  &=  0.92\pm 0.11 \pm 0.03 
  \,, \nonumber\\ 
  (m_B^2-m_\pi^2){F_+^\prime(0)\over F_+(0)} 
  &= 1.5 \pm 0.6 \pm 0.4 
  \,, \\ 
  (1-\alpha)^{-1} 
  &= 8 \,^{+2}_{-7} \pm 7
  \,. \nonumber
\end{align}
The first error is experimental, and the second is due to 
uncertainty in the form factor shape (these quantities are independent of the 
form factor normalization). 
The central values in (\ref{eq:shape}) are again obtained using the
parameterization (\ref{eq:trunc_series}) with $\sum_k a_k^2 < 0.01$, 
and the shape error is conservatively estimated by adding the maximum variation of the
boundaries of the $1\sigma$ interval  
when the bound is relaxed to $\sum_k a_k^2 < 0.1$.  

While the conventional dispersive bound approach provides an elegant means of
demonstrating formal convergence properties with the minimal
assumption of form factor analyticity and the convergence of an operator
product expansion, some caution is required in
order to avoid misinterpreting the results.  Firstly, for certain
observables, e.g. $|V_{ub}|$, 
the fits are much more tightly constrained by the data
than by the unitarity-based dispersive bound.
This leads to the happy conclusion
that the errors on $|V_{ub}|$ do not depend on 
the chosen parameterization or the exact value of the bound, 
and the analysis lends itself to a straightforward
statistical interpretation.
Secondly, other important observables,
such as the slope of the form factor, {\it are} sensitive to the
addition of more parameters than can be constrained by the data, but
are allowed by the unitarity bound.  Since this bound is
overestimated, presumably by orders of magnitude, 
a reliance on this procedure would lead to the
pessimistic conclusion that almost no information at all can be
extracted from the data for these quantities. In such cases, 
we propose to use tighter bounds, which follow from the scaling behavior
of the bounded quantity in the heavy quark limit.

Apart from establishing order-of-magnitude estimates for the bounds 
in (\ref{eq:bound_rho}) and (\ref{eq:bound})  by heavy-quark power counting, 
none of the above analysis relies on heavy-quark, large-recoil or
chiral expansions, or on the associated heavy-quark, soft-collinear or
chiral effective field theories.  However, the semileptonic data 
can be used to test predictions from these effective field theories, 
and to determine low-energy parameters that can be used as inputs
to the calculation of other processes. 
For example, using the experimental result 
${\rm Br}(B^-\rightarrow \pi^-\pi^0)=(5.5\pm0.6) \times
10^{-6}$~\cite{Aubert:2004aq} together with $|V_{ub}| F_+(0)$ from (\ref{eq:shape}), we find 
\begin{equation}\label{eq:r}
  \frac{\Gamma(B^-\rightarrow \pi^-\pi^0)}{d \Gamma(\bar B^0\rightarrow \pi^+ \ell^- \bar{\nu})/dq^2|_{q^2=0}}
  = 0.76\,^{+0.22}_{-0.18} \pm 0.05 \,{\rm GeV}^2  \,,
\end{equation}
where the first error is experimental, and the second is due to the
form factor shape uncertainty in (\ref{eq:shape}).  Such ratios
provide a strong test of factorization~\cite{Bauer:1986bm}.  The
leading-order prediction for this ratio, corresponding to the
``naive'' factorization picture where hard-scattering corrections are
neglected, yields $16\pi^2 f_\pi^2 |V_{ud}|^2 (C_1+C_2)^2/3=0.62 \pm
0.07\,{\rm GeV}^2$.  This uncertainty includes only the effects of 
varying the renormalization scale of the leading-order weak-interaction
coefficients~\cite{Buchalla:1995vs} between $m_b/2$ and $2m_b$.  This
may be compared to the prediction of Beneke and
Neubert~\cite{Beneke:2003zv} who use QCD factorization theorems for
two-body decays to work beyond leading order and include the effects
of hard-scattering terms, obtaining for the same ratio,
$0.66\,^{+0.13}_{-0.08}\,{\rm GeV}^2$.  The uncertainty in their
prediction is dominated by the uncertainty in the light-cone
distribution amplitudes (LCDAs) of the $B$- and $\pi$-mesons. Bauer et
al.~\cite{Bauer:2004tj,Arnesen:2005ez} evaluate the same factorization
theorems using a different strategy: they use experimental results for
other $B\rightarrow \pi\pi$ decays to determine the part involving the
LCDAs from data, which is possible if all power corrections, and
perturbative corrections of order $\alpha_s(m_b)$, are neglected. For
the ratio (\ref{eq:r}) they find $1.27^{+0.22}_{-0.29}\,{\rm GeV}^2$,
where we display only experimental errors.  The semileptonic data
provides important information on otherwise poorly constrained
hadronic parameters entering these processes.

As a second application, the parameter
$\delta$ measuring the relative size of hard-scattering and soft-overlap
contributions in the $B\to\pi$ form factor
can be related to the slope of the form factor at $q^2=0$~\cite{Hill:2005ju}.  
Extrapolated to zero recoil, the lattice
calculations in \cite{Okamoto:2004xg,Shigemitsu:2004ft} give
for the slope of the $F_0$ form factor, 
$\beta\equiv [(m_B^2-m_\pi)^2 F_0^\prime(0)/F_+(0)]^{-1} = 1.2\pm 0.1$. 
Together with (\ref{eq:shape}) this yields  
\begin{equation}\label{eq:delta}
\begin{aligned}
  \delta \equiv 1- {m_B^2-m_\pi^2 \over F_+(0)}\left(
    \left.\frac{dF_+}{dq^2}\right|_{q^2=0} -
    \left.\frac{dF_0}{dq^2}\right|_{q^2=0} \right) = 0.4\, \pm 0.6 \, \pm
  0.1\, \pm 0.4 \,,
\end{aligned}
\end{equation}
where the first error is experimental, the second is theoretical from the 
lattice determination of $\beta$, and the third is due to the form factor 
shape uncertainty in (\ref{eq:shape}).  
Establishing the relative size of the hard-scattering and 
soft-overlap contributions from the
semileptonic data provides another important input to
factorization analyses of hadronic $B$ decays.
The above result for $\delta$ does not unambiguously
establish $\delta\neq 0$ which signals the presence of hard-scattering
terms, but it disfavors the opposite scenario, $\delta\approx 2$, where the form
factor is completely dominated by hard-scattering.
More data will help reduce both the experimental and shape-uncertainty 
errors for this quantity. 

As a third application, the form factor $F_+(0)$ and shape observable $\alpha$ determine 
the coupling constant $g_{B^*B\pi}$ via
\begin{equation}\label{eq:coupling}
\begin{aligned}
  {f_{B^*} g_{B^*B\pi} \over 2 m_{B^*} } 
  \equiv {F_+(0)\over 1-\alpha} 
  = 2.0 \,^{+0.6}_{-1.6} \pm 0.2 \pm 1.7 \,,
\end{aligned}
\end{equation}
where the first error is experimental, the second is theoretical from
the lattice form factor normalization, and the final error is due to
the form factor shape uncertainty, determined as in (\ref{eq:shape}).
Since the semileptonic data is concentrated at small $q^2$, it is not
very sensitive to the detailed structure of the sub-threshold pole and
dispersive integral in (\ref{eq:dispersion}).  In fact, the data do
not yet definitively resolve a distinct contribution of the $B^*$
pole, although the opposite scenario --- dominance by the $B^*$ pole in
(\ref{eq:dispersion}) --- is ruled out~\cite{Hill:2005ju}.

Our implementation of the bounds in (\ref{eq:bound_rho}) and
(\ref{eq:bound}) could be formalized in terms of standard methods of
constrained curve fitting~\cite{bayesian}.  In this language, we have
enforced a ``prior'' probability function which is constant if the
parameters obey the bound on $\sum_k |\rho_k|$ or $\sum_k a_k^2$, 
and zero otherwise.  For simplicity, we then performed a
$\chi^2$ fit, assuming sufficient statistics that the data is Gaussian
distributed. The resulting error estimates should be conservative.
Firstly, this prior allows equal probability for parameter values 
that are near the bound, even though we believe 
such values are increasingly unlikely.  Other prior functions may be
considered --- for example, in the case of the series parameterization (\ref{eq:trunc_series}),  
a Gaussian prior on the variable $(-\log_{10}\sum_k a_k^2)$, 
with mean and standard deviation of order unity. 
Secondly, in estimating errors based on $\Delta\chi^2$, we neglect the
fact that bounds enforce restrictions that renormalize the probability
distributions, and to the extent that the bounds are relevant, this
tends to overestimate errors.  As a simple example, if an absolute
bound happened to coincide with the boundary of the ``$1\sigma$''
interval obtained for an observable based on $\Delta \chi^2=1$, we
would estimate that the observable was within the interval with only
$\sim 68\%$ confidence, whereas the bounds guarantee this with $100\%$
confidence.  In a more refined analysis, a direct evaluation of the
statistical integrals could account for such boundary effects.  An
alternative procedure employed in \cite{Lellouch:1995yv}, and
generalized in \cite{Fukunaga:2004zz} to include shape information
from experiment, has a slightly more complicated statistical
interpretation.  Here theory information on the form factor, combined
with the dispersive bounds, is used to generate a statistical sample
of ``envelopes'', each consisting of the curves defined at each
$q^2$-point by the minimum and maximum values that the form factor can
take.  (Note that some curves may be ruled out by the bounds, yet
allowed by the envelopes, which are generated by extremizing
point-by-point in $q^2$.)  This sample of envelopes is then combined
with experimental branching fractions to determine a distribution for
$|V_{ub}|$ or other observables.  Working in terms of parameters $a_k$
allows the experimental and lattice data to be treated on the same
footing, and yields a more straightforward interpretation of the
constraints enforced by the bounds.
Fortunately, these complications play an extremely minor role in the
case of $|V_{ub}|$.  As illustrated by Figure~\ref{fig:data}, the
errors are very nearly Gaussian, and nearly identical results are
obtained using different parameterizations, and widely different
values for the bounds.  A more refined statistical analysis might be
useful for those shape observables that show sensitivity to the
bounds, to extract as much information as possible from the
experimental data.  

In summary, we have shown that  
the form factor shape information necessary for
a precise extraction of $|V_{ub}|$ is now entirely determined from experiment. 
Rather than relying on theoretical models for this shape, the current and
future experimental data can instead be 
used as a precision tool for testing theory 
predictions and determining hadronic parameters in other processes.  
For example, the ratio in (\ref{eq:r}) should be predicted with good accuracy 
from the factorization approach to hadronic $B$ decays, and 
can be even more firmly established once the hard-scattering contribution in 
(\ref{eq:delta}) is determined more precisely from data.  
The methodology employed here in $B$ decays
can be validated in the
analogous situation of semileptonic $D$ decays, where experiment and
lattice cover the entire range of $q^2$.  
Note that we only used lattice input for the form factor at a
single $q^2$-value, 
to avoid theoretical biases on the form factor shape, and 
to emphasize the conclusion that the
shape is determined by experiment; however, studying the
form factor shape provides an important test of lattice calculations.
Our results show that with improved lattice data, an exclusive measurement 
of $|V_{ub}|$ that rivals or even surpasses the inclusive determination is
possible.

\subsection*{Acknowledgments}

We thank A.~Kronfeld, H.~Quinn and I.~Stewart for discussions.  
We are grateful to the Institute for Nuclear Theory (Seattle, WA) 
for hospitality where a part of this work was completed.  
Research supported by the Department of Energy under Grants
DE-AC02-76SF00515 and DE-AC02-76CH03000. Fermilab is operated by
Universities Research Association Inc., under contract with the U.S.\
Department of Energy.


\begin{thebibliography}{99}


\bibitem{Athar:2003yg}
  S.~B.~Athar {\it et al.}  [CLEO Collaboration],
  Phys.\ Rev.\ D {\bf 68}, 072003 (2003)
  [hep-ex/0304019].

\bibitem{Abe:2004zm}
  K.~Abe {\it et al.}  [BELLE Collaboration],
  hep-ex/0408145.

\bibitem{Aubert:2005cd}
  B.~Aubert  [BABAR Collaboration],
  hep-ex/0507003.

\bibitem{Aubert:2005tm}
  B.~Aubert  [BABAR Collaboration],
  hep-ex/0506064.

\bibitem{Okamoto:2004xg}
  M.~Okamoto {\it et al.},
  Nucl.\ Phys.\ Proc.\ Suppl.\  {\bf 140}, 461 (2005)
  [hep-lat/0409116].

\bibitem{Shigemitsu:2004ft}
J.~Shigemitsu {\it et al.},
Nucl.\ Phys.\ Proc.\ Suppl.\  {\bf 140}, 464 (2005)
[hep-lat/0408019].

\bibitem{Bourrely:1980gp}
  C.~Bourrely, B.~Machet and E.~de Rafael,
  Nucl.\ Phys.\ B {\bf 189}, 157 (1981).

\bibitem{Boyd:1994tt}
  C.~G.~Boyd, B.~Grinstein and R.~F.~Lebed,
  Phys.\ Rev.\ Lett.\  {\bf 74}, 4603 (1995)
  [hep-ph/9412324].

\bibitem{Lellouch:1995yv}
  L.~Lellouch,
  Nucl.\ Phys.\ B {\bf 479}, 353 (1996)
  [hep-ph/9509358].

\bibitem{Boyd:1997qw}
  C.~G.~Boyd and M.~J.~Savage,
  Phys.\ Rev.\ D {\bf 56}, 303 (1997)
  [hep-ph/9702300].

\bibitem{Becirevic:1999kt}
  D.~Becirevic and A.~B.~Kaidalov,
  Phys.\ Lett.\ B {\bf 478}, 417 (2000)
  [hep-ph/9904490].

\bibitem{Hill:2005ju}
  R.~J.~Hill, Phys.\ Rev.\ D (in press)  
  [hep-ph/0505129].

\bibitem{Arnesen:2005ez}
  M.~C.~Arnesen, B.~Grinstein, I.~Z.~Rothstein and I.~W.~Stewart,
  Phys.\ Rev.\ Lett.\  {\bf 95}, 071802 (2005)
  [hep-ph/0504209].

\bibitem{Boyd:1995sq}
  C.~G.~Boyd, B.~Grinstein and R.~F.~Lebed,
  Nucl.\ Phys.\ B {\bf 461}, 493 (1996)
  [hep-ph/9508211].

\bibitem{Caprini:1997mu}
  I.~Caprini, L.~Lellouch and M.~Neubert,
  Nucl.\ Phys.\ B {\bf 530}, 153 (1998)
  [hep-ph/9712417].

\bibitem{Ball:2005tb}
  P.~Ball and R.~Zwicky,
  Phys.\ Lett.\ B {\bf 625}, 225 (2005)
  [hep-ph/0507076].
 
\bibitem{Aubert:2004aq} HFAG,
  http://www.slac.stanford.edu/xorg/hfag/, using

  B.~Aubert {\it et al.}  [BABAR Collaboration],
  Phys.\ Rev.\ Lett.\  {\bf 94}, 181802 (2005)
  [hep-ex/0412037],

  Y.~Chao {\it et al.}  [Belle Collaboration],
  Phys.\ Rev.\ D {\bf 69}, 111102 (2004)
  [hep-ex/0311061],

  A.~Bornheim {\it et al.}  [CLEO Collaboration],
  Phys.\ Rev.\ D {\bf 68}, 052002 (2003)
  [hep-ex/0302026].

\bibitem{Bauer:1986bm}
  M.~Bauer, B.~Stech and M.~Wirbel,
  Z.\ Phys.\ C {\bf 34}, 103 (1987),

  J.~D.~Bjorken,
  Nucl.\ Phys.\ Proc.\ Suppl.\  {\bf 11}, 325 (1989).

\bibitem{Buchalla:1995vs}
  G.~Buchalla, A.~J.~Buras and M.~E.~Lautenbacher,
  Rev.\ Mod.\ Phys.\  {\bf 68}, 1125 (1996)
  [hep-ph/9512380].

\bibitem{Beneke:2003zv}
  M.~Beneke and M.~Neubert,
  Nucl.\ Phys.\ B {\bf 675}, 333 (2003)
  [hep-ph/0308039].

\bibitem{Bauer:2004tj}
  C.~W.~Bauer, D.~Pirjol, I.~Z.~Rothstein and I.~W.~Stewart,
  Phys.\ Rev.\ D {\bf 70}, 054015 (2004)
  [hep-ph/0401188].


\bibitem{bayesian}
  For an introduction, see 
  G.~P.~Lepage, B.~Clark, C.~T.~H.~Davies, K.~Hornbostel, P.~B.~Mackenzie, C.~Morningstar and H.~Trottier,
  Nucl.\ Phys.\ Proc.\ Suppl.\  {\bf 106}, 12 (2002)
  [hep-lat/0110175].

\bibitem{Fukunaga:2004zz}
  M.~Fukunaga and T.~Onogi,
  Phys.\ Rev.\ D {\bf 71}, 034506 (2005)
  [hep-lat/0408037].


\end{thebibliography}
\end{document}